\RequirePackage{fix-cm}
\documentclass{svjour3}                     
\smartqed  
\usepackage{graphicx}
\usepackage{mathptmx}      
\usepackage[authoryear]{natbib}
\newcommand{\eqref}[1]{(\ref{#1})} 

 \journalname{Bulletin of Mathematical Biology}
\begin{document}

\title{A model of chloroplast growth regulation in mesophyll cells
}


\author{Kelly M Paton          \and
        Lisa Anderson \and
        Pauline Flottat \and
Eric N Cytrynbaum
        }


\institute{E. Cytrynbaum \at
              Institute of Applied Mathematics, University of British Columbia \\
              Tel.: 1-604-822-3784\\
              Fax: 1-604-822-6074\\
              \email{cytryn@math.ubc.ca}           
}

\date{Received: date / Accepted: date}

\maketitle

\begin{abstract}
Chloroplasts regulate their growth to optimize photosynthesis. Quantitative data shows that the ratio of total chloroplast area to mesophyll cell area is constant across different cells within a single species, and also across species. Wild-type chloroplasts exhibit little scatter around this trend; highly irregularly-shaped mutant chloroplasts exhibit more scatter. Here we propose a model motivated by a bacterial quorum-sensing model consisting of a switch-like signalling network that turns off chloroplast growth. We calculated the dependence of the location of the relevant saddle-node bifurcation on the geometry of the chloroplasts. Our model exhibits a linear trend, with linearly growing scatter dependent on chloroplast shape, consistent with the data. When modelled chloroplasts are of a shape that grows with a constant area to volume ratio (disks, cylinders) we find a linear trend with minimal scatter. Chloroplasts with area and volume that do not grow proportionally (spheres) exhibit a linear trend with additional scatter.
\keywords{chloroplasts \and ordinary differential equations \and bifurcation \and model \and growth regulation \and switch}
\end{abstract}

\section{Introduction}\label{intro}
Chloroplasts are the organelles within plant cells which convert sunlight to usable energy via photosynthesis. Their exposure to sunlight is integral to optimal photosynthetic output, so their coverage of a cell's surface area is important. There is evidence that chloroplast growth and replication is regulated to control the percentage of a mesophyll cell's surface area that is occupied by chloroplasts. This is quantified by comparing the plan areas of the chloroplasts and the cell, defined as the areas taken up by the chloroplasts and the cell when viewed from above \citep{pyke1991rapid}. Several studies support this claim.

\begin{figure}
\includegraphics[width=\textwidth]{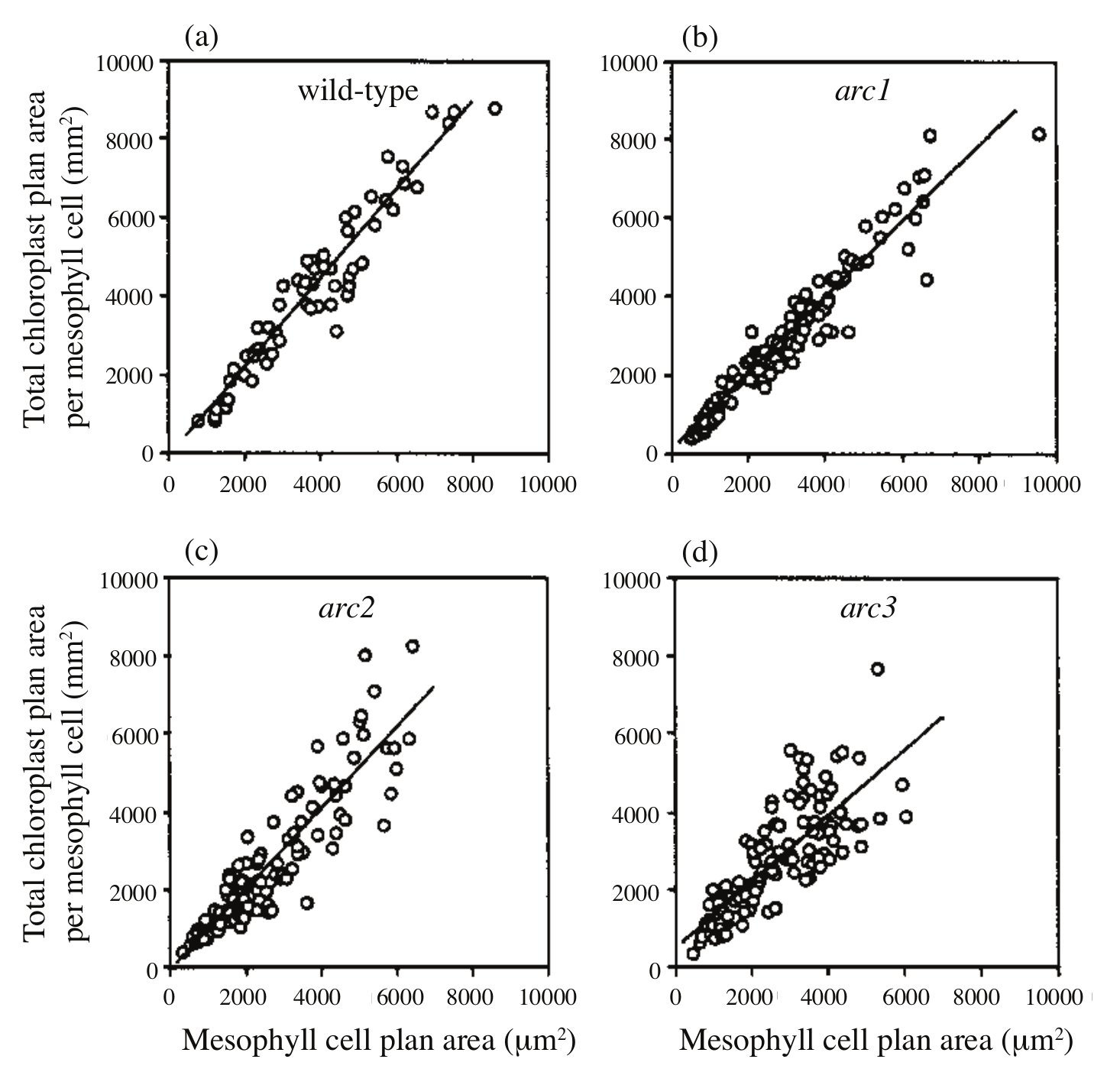}
\caption{Figure reprinted (with permission) from \cite{pyke1992chloroplast}, figure 3, showing the linear relationship between total chloroplast plan area per cell and cell plan area for each of the wild type and three mutants of \emph{Arabidopsis thaliana}. The slope of all four lines is similar. There are two layers of chloroplasts within a cell, one at the top and one at the bottom, so ``full'' cells would exhibit a slope of 2. Scatter varies, with the wild-type and \emph{arc1} mutants having the least scatter ($R^2=0.92$ and $R^2=0.93$, respectively, in (a) and (b)) and \emph{arc2} and \emph{arc3} mutants showing more scatter ($R^2=0.81$ and $R^2=0.61$, respectively, in (c) and (d)).}\label{fig:pykeLeech92-fig3}
\end{figure}

\cite{ellis1985} found that leaves of \emph{Triticum aestivum L.} and \emph{T. monocoecum L.} demonstrated a trade-off between chloroplast size and number that resulted in a similar total chloroplast plan area per unit cell plan area. \cite{pyke1992chloroplast} then examined mesophyll cells from a wild type and three chloroplast replication and growth mutants (termed ``arc" mutants) of \emph{Arabidopsis thaliana} and consistent with the observations of Ellis and Leech, they found an inverse relationship between the mean individual chloroplast area and the total number of chloroplasts (see Figure \ref{fig:pykeLeech92-fig3}). For each of the wild-type and the three mutants, the relationship between total chloroplast plan area and cell plan area was linear, no matter the individual size or shape of the chloroplasts. Moreover, the slope of the data plotted in the ``plan area plane'' (total chloroplast plan area versus cell plan area) was quantitatively similar for all four data sets. Two of the mutants showed considerably more scatter than the wild type and the other mutant, and the mutant with the most scatter was observed to have irregularly shaped chloroplasts.

\begin{figure}
\includegraphics[width=0.9\textwidth]{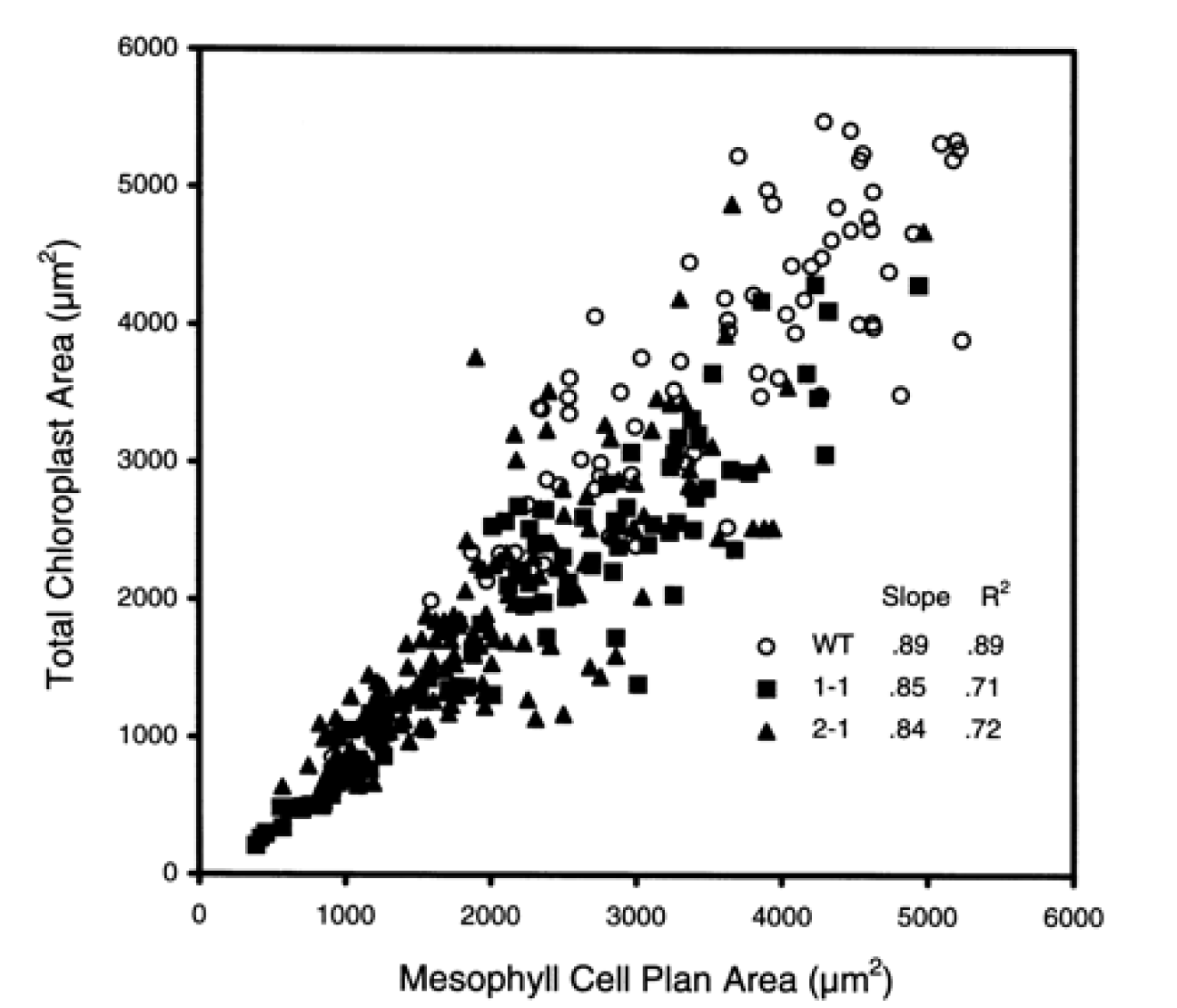}
\caption{Figure reprinted (with permission) from \cite{osteryoung1998chloroplast}, figure 5, showing the chloroplast plan areas versus cell size for wild type (WT, open circles) and two different transgenic types (1-1, filled boxes, and 2-1, filled triangles) of \emph{Arabidopsis}. All three relationships are linear with similar slopes; see calculated slopes and $R^2$ values.
}\label{fig:osteryoung-fig5}
\end{figure}

\cite{osteryoung1998chloroplast} unearthed more evidence of this phenomenon from a collection of \emph{Arabidopsis} cells of both the wild type and two transgenic types. Their figure plotting the total chloroplast plan area versus the cell plan area for each cell is duplicated here in Figure \ref{fig:osteryoung-fig5}. Similar to the data in \cite{pyke1992chloroplast}, each of the three types displayed linear relationships of similar slope, with scatter that scaled with cell size. The transgenic types -- with larger and more irregularly shaped chloroplasts than the wild type -- presented with more scatter; see the $R^2$ values in Figure \ref{fig:osteryoung-fig5}.

\cite{pyke1999plastid} further showed that this regulated chloroplast density persists not only across different cells, but across different species. He found a linear relationship between total chloroplast area and cell plan area when considering an average data point from each of eight different species and several \emph{Arabidopsis} mutants. He noted that the signalling methods used by chloroplasts to regulate their growth and division are not known, but concluded that ``chloroplast density in relation to the size of the cell seems to be involved''. 

The most notable feature in the data sets shown in Figures \ref{fig:pykeLeech92-fig3} and \ref{fig:osteryoung-fig5} is the linear relationship between total chloroplast plan area and cell plan area; this is backed up by the data from \cite{pyke1999plastid}. The second notable feature is that the scatter about the relationship grows somewhat linearly with cell plan area, more like a percentage of the plan area ratio rather than an absolute measure. This data can be resummarized in the context of the ratio of plan areas, which we'll call $\alpha$, defined as the total chloroplast plan area divided by the cell plan area. These two observations -- the linearity of the relationship and the linearly increasing scatter -- can be reduced down to the observation that the range of ratios of plan areas is constant for all cell plan areas. The third important observation is that irregularly shaped chloroplasts have more scatter, which means that chloroplast geometry matters.

Here we propose a mechanism to explain this observed chloroplast growth regulation phenomenon. The ability to sense the collective density of one's own population is found in certain bacteria and is referred to as ``quorum sensing'' \citep{dockery2001}. Much work has been focused on understanding quorum sensing; \cite{dockery2001} modelled the biochemical quorum-sensing switch in \emph{Pseudomonas Aeruginosa}.

At the core of our model is the Dockery-Keener biochemical switch. In bacteria, flipping this switch triggers a shift in gene regulation allowing a sufficiently large population to collectively change behaviour. These triggered behaviours include producing exopolysaccharide to form a biofilm, secreting toxins and turning on bioluminescence. In chloroplasts, we propose that a similar switch turns off chloroplast growth.

Although chloroplasts may have a quorum-sensing biochemical pathway homologous to that of bacteria, we prefer to think of this quorum-sensing model simply as a biochemical system with switch-like behaviour. This switch-like behaviour is the important feature for our results. Even if the chloroplast growth regulation pathway operates differently, any switch-like behaviour involving secretion of the signal ought to behave similarly to what we describe here. We find that the model, when applied to a system of chloroplasts within a cell, can capture both the linear trend and the linearly growing scatter in the data. In the model, chloroplast shape influences the extent of the scatter.

\section{Mathematical model}\label{sec:model}

\subsection{Cell and chloroplast geometry}\label{sec:geom}

Quorum sensing is dependent on the geometry of the system. In a system comprised of chloroplasts within a leaf cell, the biochemical processes of secretion, degradation, and absorption are coupled to geometry via dependence on the surface area of the chloroplasts and the volume density of the chloroplasts within the cell. We define the geometry of the system in order to find explicit definitions of the chloroplast density, $\rho$, and the chloroplasts' effective membrane permeability, $\delta$ -- which is dependent on surface area -- to use in the quorum-sensing model.

\begin{figure}
\includegraphics[width=0.6\textwidth]{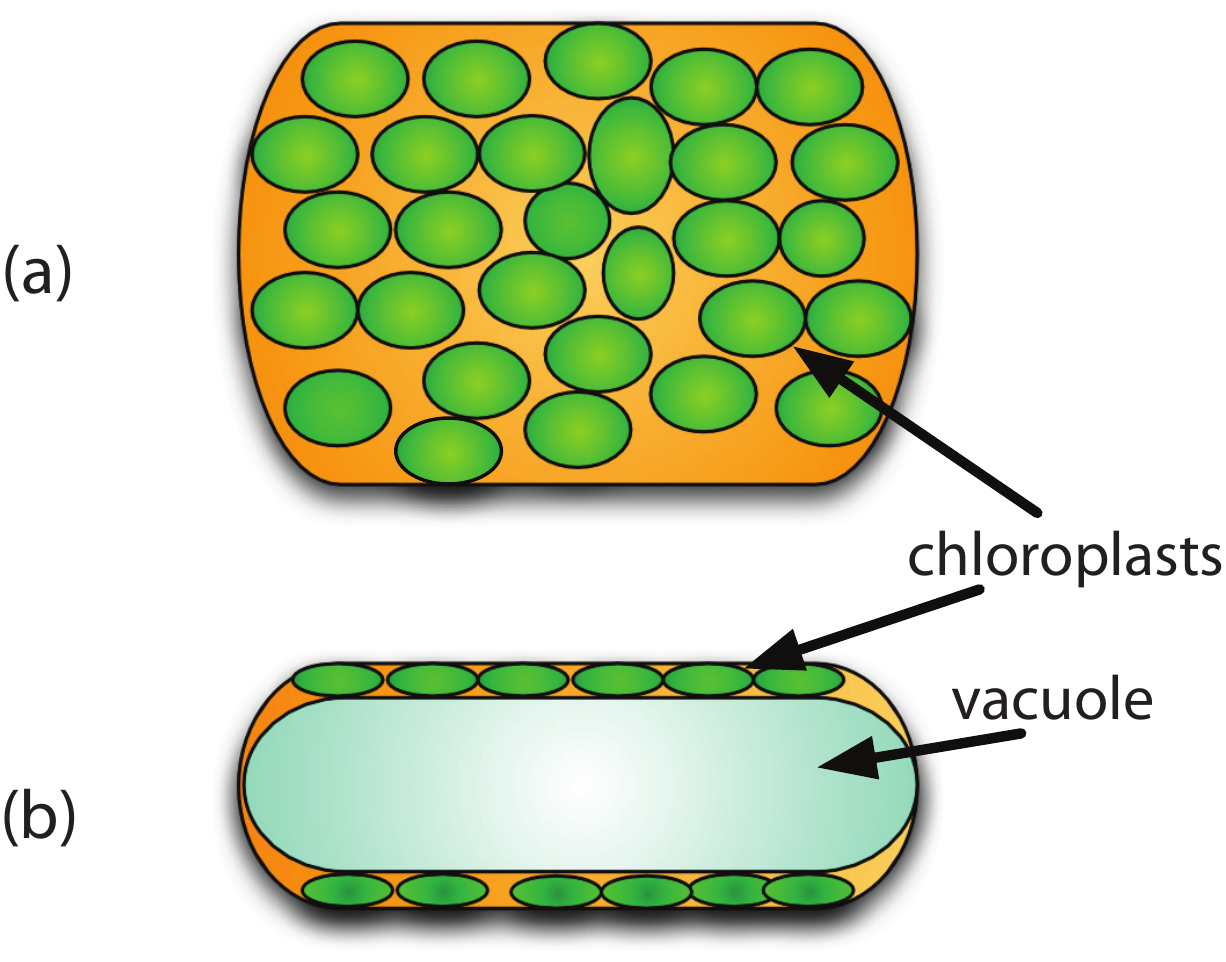}
\caption{This cartoon summarizes the features of \emph{A. thaliana} leaf cell images (see examples in \cite{pyke1994arc6} and \cite{hall1996molecular}) from both (a) an above view and (b) a transverse section view. The large vacuole occupying the middle of the cell pushes the chloroplasts into a thin layer against the cell wall, usually concentrated on the top and bottom for better access to light sources. Based on this cartoon, we model the relevant cytosolic space as two thin layers of thickness $\tau$ and top surface area $S$ that can each hold a single layer of chloroplasts, one at the top and one at the bottom. Figure created in OmniGraffle.}\label{fig:cellCartoon}
\end{figure}

Consider a cell of fixed size that contains a population of chloroplasts. A leaf mesophyll cell of \emph{A.\ thaliana} is largely occupied by one or more vacuoles, leaving only a thin layer of the cytosol available for the chloroplasts at the surface \citep{pykeplastidbook, pyke1999plastid}; see Figure \ref{fig:cellCartoon}. We model this available cytosolic space as two thin slices that can each hold a single layer of chloroplasts, with a top surface area of $S$ and a thickness of $\tau$. The volume of a cell is thus $V=2S\tau$, and its plan area (area viewed from above) is $S$. We look at a range of cell sizes by considering a range of $S$ values; we keep the thickness of each cytosolic space constant at $\tau$. 

\begin{figure}
\includegraphics[width=\textwidth]{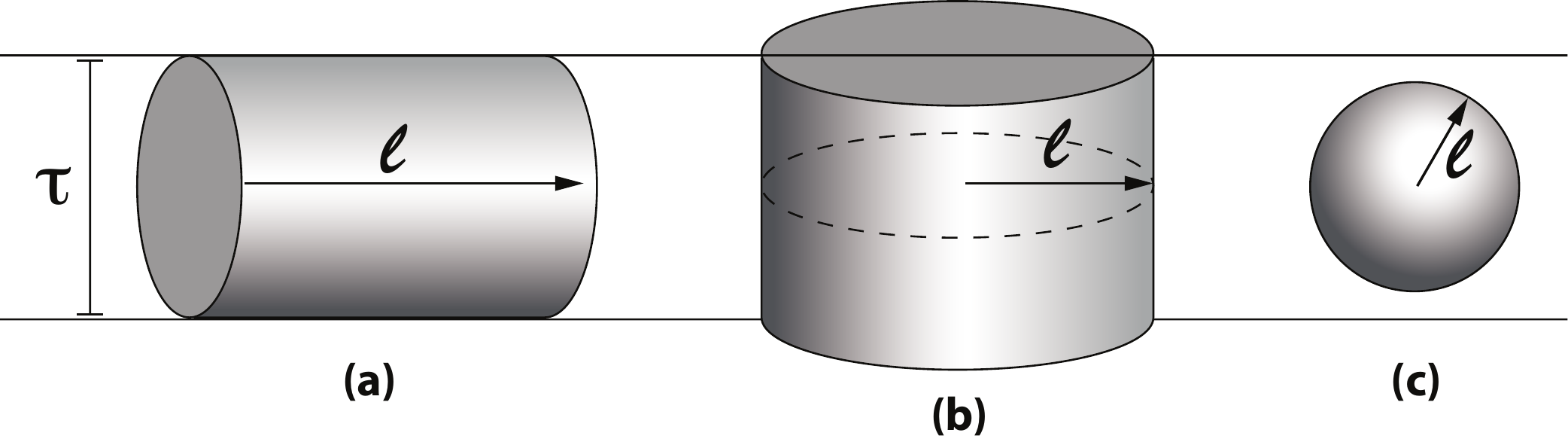}
\caption{Chloroplasts are modelled either as all cylinders as in (a); all disks as in (b); or all spheres as in (c). Orientation is as shown, within a layer of fixed thickness $\tau$. Each shape grows in the direction labelled $\ell$; all other dimensions are fixed. Our model contains two such layers, of top surface area $S$. Figure created in Adobe Illustrator.}\label{fig:chloroshapes}
\end{figure}

Chloroplast shape is more complicated. Chloroplasts typically resemble flattened footballs, although this varies widely \citep{pykeplastidbook}. The variation in their morphology is both characteristic of different plant species or cell types, and dynamic in response to the environment. For example, the \emph{arc6} mutant of \emph{A. thaliana} has highly irregularly shaped chloroplasts in comparison to the wild-type chloroplasts \citep{pyke1994genetic}. Dynamically, chloroplasts can deform in response to being squeezed against neighbouring plastids or vacuoles or the cell wall, which can happen over long time scales as the cell grows, or in shorter time scales as chloroplasts move in response to changes in light \citep{pykeplastidbook, briggs2002}. Chloroplast morphology also changes during division when the dividing chloroplast goes through a ``dumbell-shaped" stage \citep{pykeplastidbook}. Since there is such variability in chloroplast shape, we represent chloroplasts as either all cylinders, all thin disks, or all spheres, and we fix all the dimensions except for one which we refer to as the growth direction, $\ell$. Cylinders are of a fixed radius $\tau/2$ but variable length $\ell$; disks are a fixed thickness $\tau$ with variable radius $\ell$; spheres have a variable radius $\ell<\tau/2$. See Figure \ref{fig:chloroshapes}.

For any of these three basic chloroplast shapes, we can define the surface area of each chloroplast as $S_c$, and the volume of each chloroplast as $V_c$. In a cell of volume $V=2S\tau$, the chloroplast density is then $\rho=n V_c/2S\tau$ and the effective permeability is $\delta=n S_c \tilde{\delta}$, where $n$ is the number of chloroplasts and $\tilde{\delta}$ is the per surface area permeability of the chloroplast membrane.

\subsection{Quorum sensing equations}\label{sec:qs}

In this section, we reformalize the Dockery-Keener model to make it suitable for the chloroplast context by explicitly incorporating the parameters that govern cell and chloroplast geometry ($\rho, \delta$).

\cite{dockery2001} based their quorum sensing model on the eight component gene-regulatory system described in \cite{van1998cell}. They simplified it down to two ordinary differential equations using time scale assumptions. We adopt the two variable model in this paper but consider chloroplasts within a cell in place of cells within a matrix, and chloroplast density instead of cell density. The first equation tracks the concentration of an autoinducer ($A$), a signalling molecule which is secreted and can be sensed by all chloroplasts. The second equation tracks the concentration of a protein ($R$) that dimerizes with the autoinducer to form an activator of autoinducer production, thereby generating a positive feedback loop. When the autoinducer reaches a critical level, we assume (but do not explicitly model) that a downstream pathway is triggered and chloroplast growth is shut off. The equations for $A$ and $R$ are

\begin{equation}
\frac{dR}{dt}=V_{R}\frac{RA}{K_{R}+RA}+R_{0}-k_{R}R,\label{eq:dR} 
\end{equation}

\begin{equation}
\frac{dA}{dt}=V_{A}\frac{RA}{K_{A}+RA}+A_{0}-dA,\label{eq:dA}
\end{equation}

where 
\begin{equation}
d=k_{A} + \frac{\delta}{\rho}\frac{k_E(1-\rho)}{k_{E}(1-\rho)+\delta}.
\end{equation}

Here $k_E$ and $k_A$ are degradation rates of the signalling molecule outside and inside the chloroplast, respectively; $k_R$ is the degradation rate of the protein $R$ inside the chloroplast; $V_R$, $V_A$ control the maximum rate of Michaelis-Menten-type production, and $K_A$, $K_R$ control the shape of the production curve. $A_0$ and $R_0$ are background production rates. The parameter $d$ depends on $\rho$, the relative volume density of chloroplasts (with maximum 1), and $\delta$, the effective permeability of the chloroplast membrane. This complicated expression arises from the reduction of the system from eight variables down to two. It consists of two terms. The first term is the degradation rate constant for autoinducer inside the chloroplast. The second term gives the secretion rate when the cytosolic concentration of autoinducer is assumed to be in quasi-steady state. We use $d$ as the parameter for our bifurcation analysis because it is the aggregate parameter through which the system depends on chloroplast geometry ($\delta$) and population density ($\rho$). Using the geometry defined in Section \ref{sec:geom}, the bifurcation parameter $d$ can be written as a function of $\ell$ and $n$ for each fixed cell surface $S$ (denoted by a subscript $S$):

\begin{equation}
d_S(\ell, n)=k_{A}+\frac{2S\tau S_c(\ell) \tilde{\delta}}{V_c(\ell)}\frac{k_E\left(1-\frac{n V_c(\ell)}{2S\tau}\right)}{k_{E}\left(1-\frac{n V_c(\ell)}{2S\tau}\right)+n S_c(\ell) \tilde{\delta}},\end{equation}
where $S_c$ and $V_c$ will vary depending on the shape and size of the model chloroplasts. When $d$ is written in this form it becomes clear that chloroplast geometry and growth plays an important role in the dynamics of this system.

\begin{figure}
\includegraphics[width=0.8\textwidth]{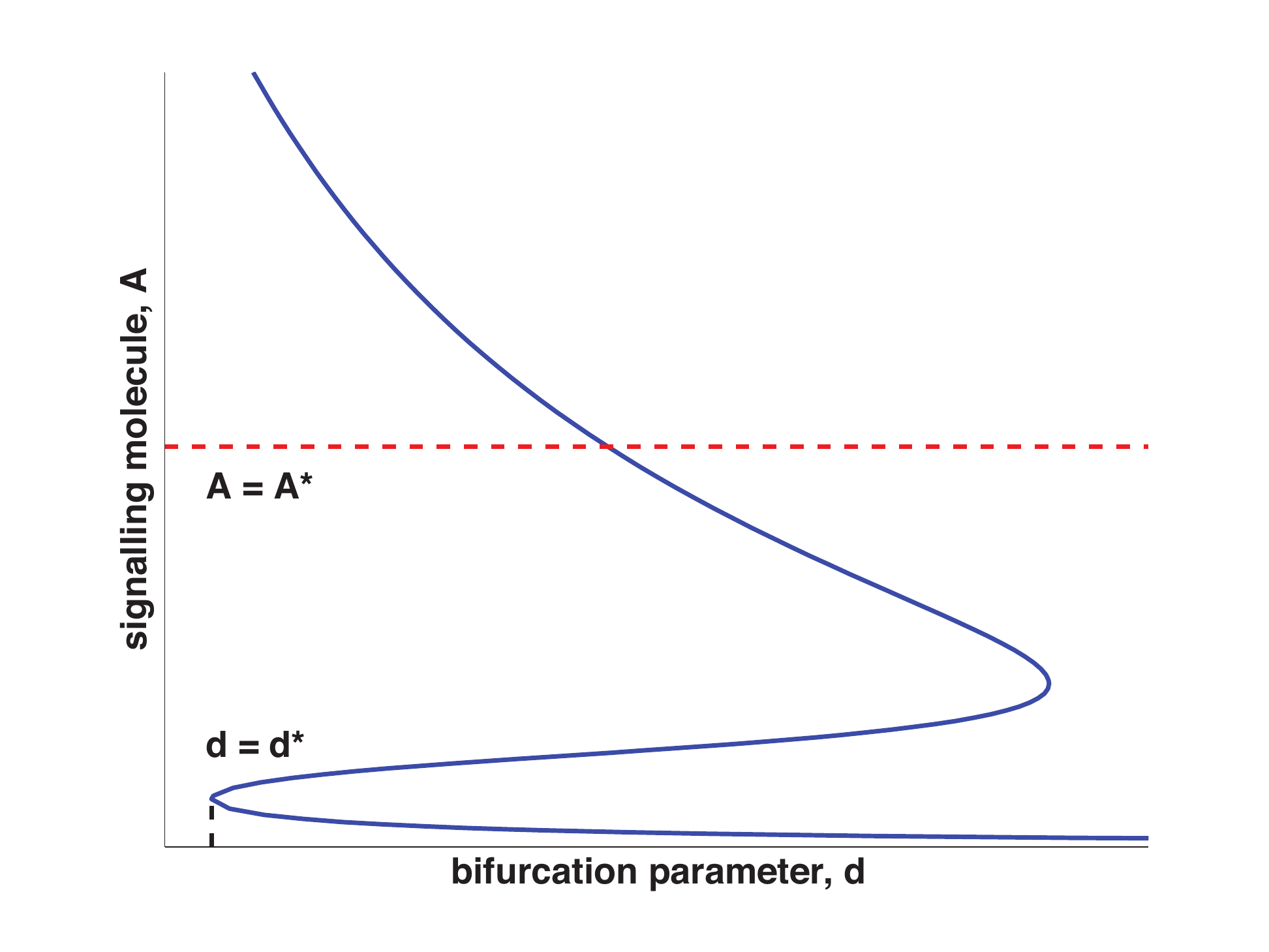}
\caption{The solid curve is a bifurcation plot showing the concentration of signalling molecule $A$ versus the bifurcation parameter $d$. A cell with low chloroplast density will begin on the right side of the lower branch and move left ($d$ will decrease) as the chloroplasts grow. When $d$ reaches the bifurcation point at $d^*$, the system will fall off the knee and the concentration of signalling molecule $A$ will rise above the critical value $A^*$ (indicated by a horizontal dashed line). This transition is the end-of-growth bifurcation at which the chloroplasts will stop growing. The bifurcation diagram was calculated using AUTO and plotted using MATLAB.}\label{fig:AUTOd}
\end{figure}

\begin{figure}
\includegraphics[width=0.8\textwidth]{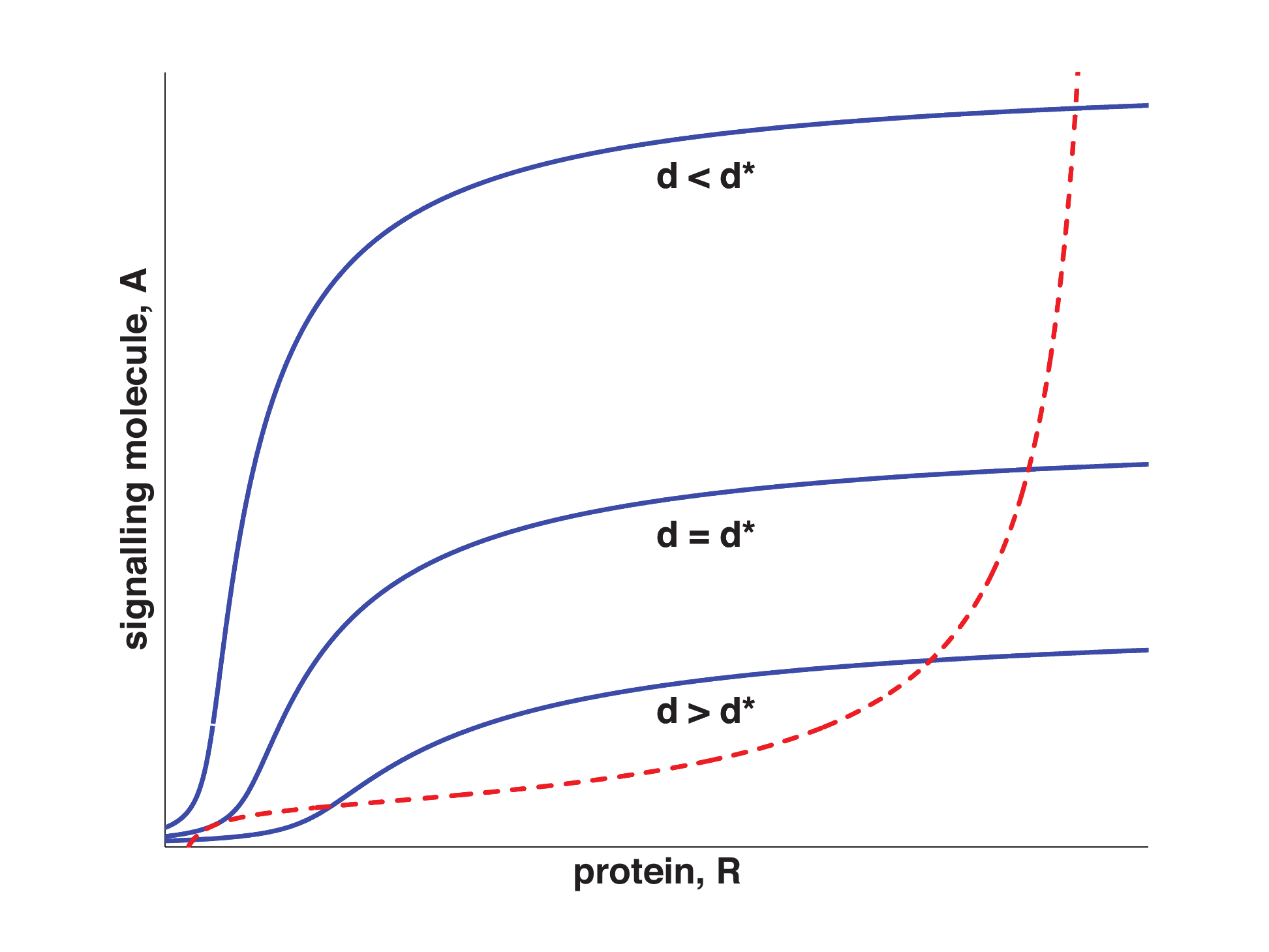}
\caption{The nullclines in the phase plane of the signalling molecule $A$ vs the protein $R$, plotted for three values of the bifurcation parameter $d$, show the passage through the first saddle-node bifurcation (the end-of-growth bifurcation). As chloroplast density increases in a cell of fixed size, $d$ will begin above $d^*$ and decrease through the bifurcation point at $d^*$. The curve $A'=0$ is solid; $R'=0$ is dashed, and is invariant to changes in $d$. Figure created in MATLAB.}\label{fig:AVSR}
\end{figure}

The steady states are found by solving a cubic in $RA$. Figure \ref{fig:AUTOd} shows a bifurcation plot of the steady state value of $A$ versus $d$. Each of the two ``knees" indicates a saddle-node bifurcation. As the chloroplasts grow, $d$ decreases, so the biologically relevant bifurcation is the one at the lower $d$ value; we label this $d$-value as $d^*$. A cell with low chloroplast density $\rho$ will begin on the lower branch at the right-hand side (the low steady state) and then track left as the chloroplasts grow, falling off the knee onto the upper branch (the high steady state). This bifurcation -- depicted in the phase plane in Figure \ref{fig:AVSR} -- is a biochemical switch.

We assume that the chloroplasts' growth is binary (on or off) dependent on the magnitude of $A$. While $A$ is below some critical value $A^*$, they remain in the growth regime. When $A$ rises above $A^*$, the chloroplasts stop growing. We refer to this switch as the end-of-growth bifurcation.

We are interested in determining the relationship between size, number, and shape of chloroplasts at the critical $d^*$ value, and how well that relationship predicts the data presented by \cite{pyke1992chloroplast} and \cite{osteryoung1998chloroplast}. In particular we aim to determine how the linear trend and linearly growing scatter of the data, and the variance in the amount of scatter, depend on chloroplast geometry.

\section{Analysis}

\subsection{General procedure}

The following analysis is performed for each of the three chloroplast geometries.

We fix cell size $S$ and denote dependence on cell size by a subscript $S$ where relevant. Since the chloroplasts vary in only one dimension the bifurcation parameter $d_S$ is a function of only two variables: the growth dimension, $\ell$, and the number of chloroplasts $n$. The form varies depending on which geometry we consider. Our goal is to find which combinations of size and chloroplast number correspond to the bifurcation point at $d_S=d^*$, and then calculate the corresponding plan areas of those size--number combinations. Using those data points we can then draw the plan area plane (total chloroplast plan area versus cell plan area $S$) for each case.

Finding the bifurcation point requires solving the equation $d_S(\ell,n)=d^*$. We nondimensionalize this equation and reformulate it as a search for the zeros of the function $g_S(\ell,n)$ defined by
\begin{equation}g_S(\ell,n)=(D^*-K)-\frac{S_c(\ell)\tau}{V_c(\ell)}\frac{\left(1-\frac{n V_c(\ell)}{2S\tau}\right)}{\beta_S\left(1-\frac{n V_c(\ell)}{2S\tau}\right)+\frac{n S_c(\ell)}{S}},\label{eq:g=0}\end{equation}
where $D^*$ is the nondimensional bifurcation point defined as $D^*=d^*/k_E$, $\beta_S=k_E/S\tilde{\delta}$, and $K=k_A/k_E$. Each $(\ell,n)$ pair corresponding to the bifurcation point describes a size $\ell$ and number $n$ of chloroplasts at which growth shuts off. The allowable pairs are restricted by the physical requirement that the chloroplasts must fit within the cell: $\frac{n V_c(\ell)}{2S\tau}<1$. Note that $n$ is discrete, so the bifurcation ``curve'' in the $\ell - n$ plane consists of a collection of discrete points.

Once we have a collection of these $(\ell,n)$ pairs that correspond to the bifurcation curve, we can calculate the ratio of the total chloroplast plan area to the cell plan area -- the slope of the line through the origin and data point in the plan area plane -- for each pair. We label this ratio generally as $\alpha_S(\ell,n)$; an explicit definition depends on the chloroplast geometry that we're working with. We then plot the points $(S, \alpha_S \cdot S)$ in the plan area plane. 

We repeat these calculations over a range of fixed $S$ values to generate the full data set.

The scatter and linearity of the data can also be examined analytically for each geometric case by looking at the expressions for $\alpha_S$; we include this below on a case by case basis. The range of $\alpha_S$ values dictates the vertical scatter of the data at that fixed $S$-value on the x-axis. It is possible that different $(\ell,n)$ pairs return the same $\alpha_S$ value so the vertical scatter does not necessarily correlate to the number of bifurcation points. The variance in the range of $\alpha_S$ values across the different cell sizes $S$ dictates the shape or trend of the data as a whole. If the range of values of $\alpha_S$ is invariant in $S$ (i.e., $\alpha_S=\alpha$), then the predicted chloroplast plan area to cell plan area relationship is linear with linearly growing scatter.

\subsection{Cylinder-shaped chloroplasts}

A right cylinder of variable length $\ell$ and fixed diameter $\tau$ has a surface area of $S_c=\pi \tau^2/2 + \pi \tau \ell$ and volume $V_c=\pi \tau^2 \ell/4$. To keep the chloroplast dimensions in a physically reasonable regime we consider $\ell>\tau/2$. When the chloroplasts take this shape in a cell of fixed surface area $S$ and thickness $\tau$, the function in (\ref{eq:g=0}) becomes
\begin{equation}
g_S(\ell,n)=(D^*-K)-{8\left( \frac{\tau}{2\ell}+  1\right)}\frac{\left(1-\frac{n \pi {\tau} \ell}{8S}\right)}{\beta_S\left(1-\frac{\pi {\tau} \ell n}{8S}\right)+\frac{\pi \tau \ell n}{S} (\frac{\tau}{2\ell} + 1)}.\label{eq:gcyl}
\end{equation}
To find the roots -- the $(\ell,n)$ pairs that correspond to the end-of-growth bifurcation for this $S$ -- we set $g_S(\ell,n)=0$. This can be solved numerically using Newton's method (for the general case), or in this case analytically as a quadratic equation in $\ell$ (taking only the positive root).
Then the plan area ratio for cylinders, defined as
\begin{equation}
\alpha_{S_{cylinder}} = \frac{\tau \ell n}{S},\label{eq:acylinder}
\end{equation}
is calculated for each $(\ell,n)$ pair at the end-of-growth bifurcation.
Each unique $\alpha_{S_{cylinder}}$ value corresponds to a point $(S, \alpha \cdot S)$ in the plan area plane. The range of $\alpha\cdot S$ values provides the vertical distribution of the data at that cell size $S$. 

\begin{figure}
\includegraphics[width=0.9\textwidth]{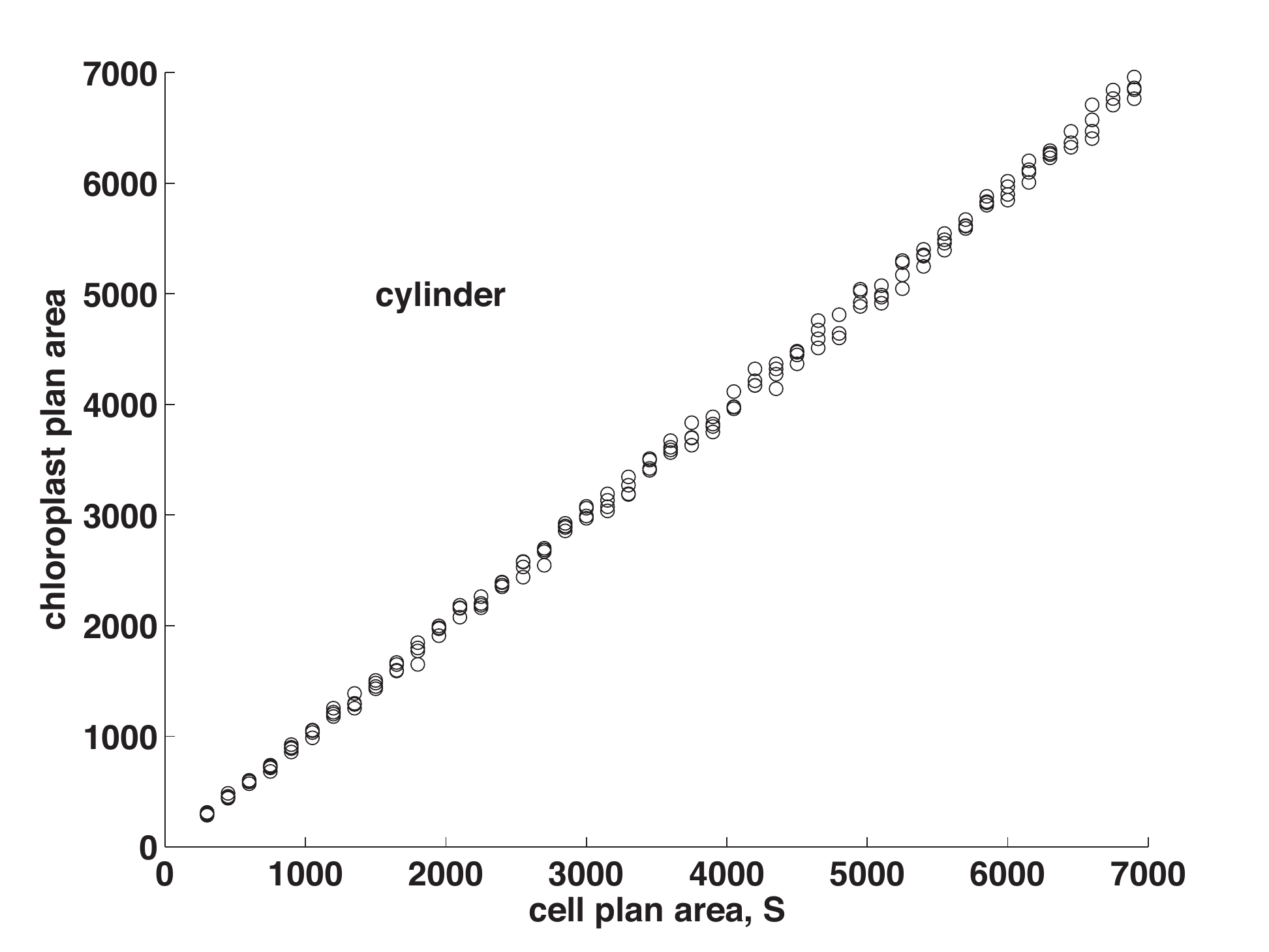}
\caption{\small{A sample of calculated cylindrical chloroplast total plan areas that correspond to the end-of-growth bifurcation, over a range of cell plan areas. Note the linear trend, with little scatter. Areas are in $\mu m^2$. Figure created in MATLAB. [Parameters: $\tilde{\delta}=5$, $k_E=0.1$, $k_A=0.4$.]}}\label{fig:cylinder}
\end{figure}

We repeat this process for a range of $S$ values, generating a set of data points at each $S$. A sample plot showing the scatter of plan areas for the full distribution of predicted chloroplast size-number pairs for cylinder-shaped chloroplasts at the end-of-growth bifurcation is shown in Figure \ref{fig:cylinder}. The apparent slope of the model data $(S, \alpha \cdot S)$ depends on as yet unconstrained parameter values. In the figures, we have chosen $k_A$ so as to get a slope close to what is observed in mesophyll cells. All parameter choices are listed in Appendix \ref{params}.

To gain further insight into the scatter and linear trend of this distribution, we solve \eqref{eq:acylinder} for $n$ and replace $n$ in $g_S=0$ from \eqref{eq:gcyl}, rearranging to find an expression for $\alpha_{S_{cylinder}}$ in terms of $\ell$:
\begin{equation}
\alpha_{S_{cylinder}}=\frac{8}{\pi}\frac{8(1+\frac{\tau}{2\ell}) - \beta_S (D^*-K)}{8(1+\frac{\tau}{2\ell})(D^*-K+1)-\beta_S(D^*-K)}\label{eq:acylinderfull}
\end{equation}
This is subject to the restrictions $D^*-K>0$ and $\beta_S>0$, and we require $\alpha>0$. To examine the scatter of points at each $S$ value, consider varying $\ell$. Under the domain constraint $\ell>\tau/2$, this expression has little variability as a function of $\ell$ and thus only a small range of $\alpha$ values for each $S$. This translates to the small amount of scatter that we see in Figure \ref{fig:cylinder}. 
Next, to examine the overall trend of the data as cell size $S$ varies, we note that the only $S$-dependence is in $\beta_S=k_E/S\tilde{\delta}$. Since $S$ is large with respect to $k_E/\tilde{\delta}$, $\beta_S$ is small for any $S$. This means that there is little variance in $\alpha_{S_{cylinder}}$ with respect to $S$, which explains the linear trend seen in Figure \ref{fig:cylinder}. 
To see both of these observations more clearly, we use the fact that $\beta_S$ is small to express $\alpha_{S_{cylinder}}$ to leading order in $\beta_S$ as
\begin{equation}
\alpha_{S_{cylinder}}\sim \frac{8}{\pi(D^*-K+1)}.\label{eq:asymacylinder}
\end{equation}
This leading order $\alpha_{S_{cylinder}}$ is constant in both $\ell$ and $S$, which predicts a linear trend with linearly growing but minimal scatter. These analytical observations match what we observe in our calculated data in Figure \ref{fig:cylinder}.

\subsection{Disk-shaped and spherical chloroplasts}

\begin{figure}
\includegraphics[width=0.9\textwidth]{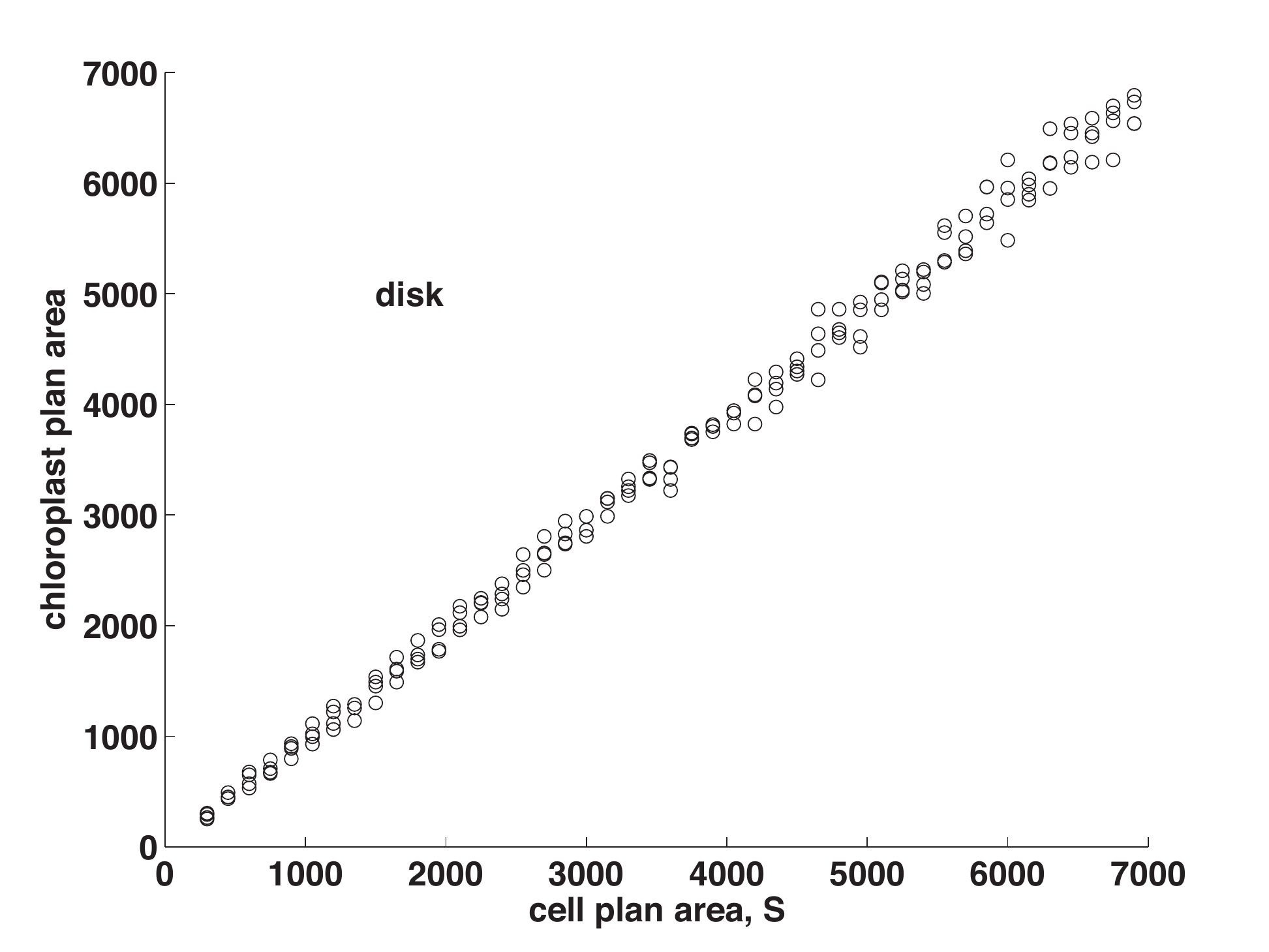}
\caption{\small{A sample of calculated disk-shaped chloroplast total plan areas that correspond to the end-of-growth bifurcation, over a range of cell plan areas. Note the linear trend, with some scatter. Areas are in $\mu m^2$. Figure created in MATLAB. [Parameters: $\tilde{\delta}=5$, $k_E=0.1$, $k_A=0.45$.]}}\label{fig:disk}
\end{figure}

\begin{figure}
\includegraphics[width=0.9\textwidth]{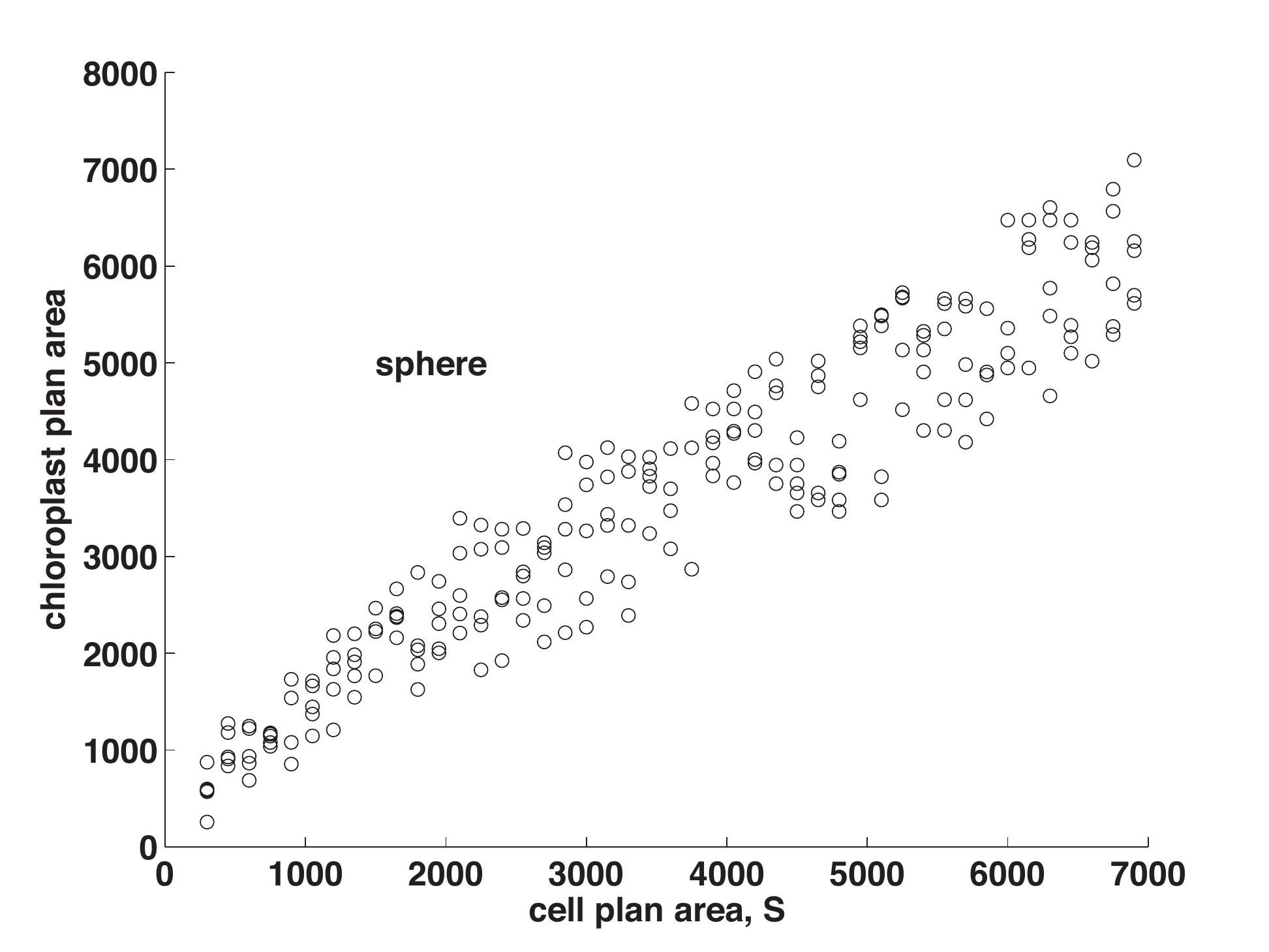}
\caption{\small{A sample of calculated spherical chloroplast total plan areas that correspond to the end-of-growth bifurcation, over a range of cell plan areas. Note the linear trend, with significant scatter. Areas are in $\mu m^2$. Figure created in MATLAB. [Parameters: $\tilde{\delta}=5$, $k_E=0.1$, $k_A=0.25$.]}}\label{fig:sphere}
\end{figure}

We perform the above analysis for disk-shaped chloroplasts, with a fixed thickness of $\tau$ and a variable radius of $\ell$, and spherical chloroplasts with a variable radius $\ell$ restricted to be $\ell<\tau/2$. The disk shape has a surface area of $S_c = 2 \pi \ell (\ell + \tau)$ and a volume $V_c = \pi \ell^2 \tau$; the sphere has surface area $S_c = 4 \pi \ell^2$ and volume $V_c = \frac{4}{3}\pi \ell^3$. Both disks and spheres look the same from above, resulting in identical plan area ratio definitions:
\begin{equation}
\alpha_{S_{disk}}= \frac{\pi \ell ^2 n}{S}, \quad
\alpha_{S_{sphere}}= \frac{\pi \ell ^2 n }{S}.\label{eq:alphasphere}
\end{equation}
Again, we explicitly calculate the $(\ell,n)$ pairs that correspond to the bifurcation point by finding the roots of \eqref{eq:g=0} with the appropriate $S_c(\ell)$ and $V_c(\ell)$ expressions, and then calculate the $\alpha_S$ values for each $(\ell,n)$ pair. We repeat this for each cell size $S$, plotting all $(S,\alpha_S \cdot S)$ points in the plan area plane. See this calculated data for disk-shaped chloroplasts in Figure \ref{fig:disk} and spherical chloroplasts in Figure \ref{fig:sphere}.

To explain the observed scatter and the linear trend in these figures we look at the expressions for $\alpha_S$. As with the cylinders, we use the definition of $\alpha_S$ in each case \eqref{eq:alphasphere} and each geometry's version of $g_S=0$ from \eqref{eq:g=0} to write $\alpha_S$ as a function of only $\ell$, analogous to \eqref{eq:acylinderfull}. 

The calculated expression for $\alpha$ as a function of $\ell$ for the disk is
\begin{equation}
\alpha_{S_{disk}}=2\frac{4(1+\frac{\tau}{\ell}) - \beta_S (D^*-K)}{4(1+\frac{\tau}{\ell})(D^*-K+1)-\beta_S(D^*-K)}.
\end{equation}
The disk expression has slightly more variance with $\ell$ than the cylinder due to the factor of $(1+\frac{\tau}{\ell})$ instead of the cylinder's $(1+\frac{\tau}{2\ell})$; this shows up as slightly more scatter in the data. The scatter is still minimal, though, which is evidenced by the leading order expression
\begin{equation}
\alpha_{S_{disk}}\sim \frac{2}{(D^*-K+1)},\label{eq:adiskasym}
\end{equation}
which is independent of $\ell$. As in the cylinder case, the scatter in the disk case is also restricted to the higher order terms. The sphere expression, on the other hand, is
\begin{equation}
\alpha_{S_{sphere}}=\frac{3\tau}{2\ell}\frac{6\frac{\tau}{\ell} - \beta_S (D^*-K)}{6\frac{\tau}{\ell}(D^*-K+1)-\beta_S(D^*-K)}
\end{equation}
which is similar in form to the other two but multiplied by $\frac{\tau}{\ell}$. That factor appears due to the way the $\delta$ and $\rho$ factors simplify. It translates to a larger range of $\alpha_S$ values for each cell size $S$ due to the stronger dependence on $\ell$, which means that the spherical chloroplasts exhibit considerably more scatter than the cylindrical or disk-shaped chloroplasts. 

The leading-order expression for the sphere $\alpha_S$ is
\begin{equation}
\alpha_{S_{sphere}}\sim \frac{3\tau}{2\ell}\frac{1}{(D^*-K+1)},\label{eq:asphereasym}\end{equation}
from which we clearly see that there is $\ell$-dependent scatter even at leading order. However, $\alpha_{S_{sphere}}$ is still independent of $S$ at leading order, which agrees with the linear trend of the data.

\section{Discussion}

Here we've proposed a model to understand the linear trend and scatter seen in the data from \cite{pyke1992chloroplast} and \cite{osteryoung1998chloroplast}. Our model explains both the linearity and scatter as being dependent on the geometry of growth. For cylinders or disks that grow in one dimension, our model predicts a linear trend with minimal scatter. For spheres, we see more scatter. This is consistent with mutant phenotypes that are characterized by irregular shapes and greater scatter when compared to the more cylindrical wild-type chloroplasts.

\subsection{Geometry effects}

If the geometry of a chloroplast has a surface area, plan area, and volume that scale together as it grows, at the end-of-growth bifurcation we get a narrow range of plan area ratios $\alpha$ for each cell size for any reasonable range of parameter values. For example, the cylindrical- and disk-shaped chloroplasts, whose surface area, plan area, and volume all scale with $\ell$ and $\ell^2$, respectively, show minimal scatter in the model. If the areas and volume do not scale together, then we get a wide range of plan area ratios for each cell size, which shows up as scatter. This appears in the case of spherical chloroplasts, which have a volume that scales with $\ell^3$ but surface and plan areas that scale with $\ell^2$. In all cases, whether the range of $\alpha$ values is small or large, the actual $\alpha$ values have little dependence on cell size $S$. As a result we observe a linear relationship between total chloroplast plan area and cell plan area, with linearly growing scatter.

\subsection{Data comparison}

The $R^2$ values for the data presented by \cite{pyke1992chloroplast} indicate that although all the data sets have a linear trend, the amount of scatter varies amongst the wild-type and the mutants (see Figure \ref{fig:pykeLeech92-fig3}). Our model can explain the variance in scatter by considering geometry. The data sets with little scatter -- the wild type, with $R^2=0.92$, and the \emph{arc1} mutant with $R^2=0.93$ -- can be explained by chloroplasts that have areas and volume that scale together as they grow, such as the cylinders or disks examined here. The data sets with additional scatter -- the \emph{arc2} mutant, with $R^2=0.81$, and the \emph{arc3} mutant with $R^2=0.61$ -- can be explained by chloroplasts that have areas and volumes that scale disproportionally, such as spheres. All three shapes considered here produce linear trends as seen in the data. 

Revisiting the original data and measuring either the number of chloroplasts in each cell or the size of the chloroplasts, for each data point, would allow us to predict where those points should fall in the plan area plane based on our model.  A comparison of the residuals for a least squares fit could tell us whether our model accounts better for the scatter in the data than noise would.

The features of the data from \cite{osteryoung1998chloroplast} are comparable to those from \cite{pyke1992chloroplast}, and can be similarly explained by our model if the transgenic types with higher scatter have chloroplast shapes with volume and area that scale disproportionally.

\subsection{Model feasibility}

Our model proposes that chloroplasts regulate their growth via a geometry-dependent biochemical switch. This idea came directly out of the bacterial quorum-sensing literature without justification. 
We do note that the endosymbionic theory traces the origins of chloroplasts and mitochondria to bacteria \citep{martin1998gene}. Moreover, \cite{pyke1999plastid} comments that ``the discovery of bacterial cell division gene homologs strongly suggests that higher plant plastid division is based on a system that has evolved from that utilized in prokaryotic cells," leaving open the possibility that plastid (i.e., chloroplast) density regulation could also have evolved from a bacterial quorum-sensing system.

That possibility aside, our model suggests constraints on what features are required of a system similar to what we've proposed: a biochemically-triggered growth switch, and some chloroplast-surface-area-dependent secreted signal which is produced proportional to chloroplast volume. This latter requirement could be complicated depending on where the genes reside (nuclear or chloroplastular).

\appendix

\section{Parameter choices}\label{params}

The ODE system constants in equations \eqref{eq:dR} and \eqref{eq:dA} were taken directly from \cite{dockery2001}: $V_R=2$, $V_A=2$, $K_R=1$, $K_A=1$, $R_0=0.05$, $A_0=0.05$, $k_R=0.7$. With these constants, the relevant bifurcation value of $d$ is at $d^*= 0.5568$. Units of time and concentration have been implicitly nondimensionalized. 

The parameters of cell thickness $\tau$ and surface area $S$, rates $k_A$ and $k_E$, and permeability $\tilde{\delta}$, that determine the value of the bifurcation parameter $d$ are chosen as follows. 

Cell size $S$ is variable but of order ${10^{3}}$ $\mu m^2$ (see Figure \ref{fig:pykeLeech92-fig3} as duplicated from \cite{pyke1992chloroplast}) and cell thickness $\tau$ is constant and of order $1$ $\mu m$ \citep{pyke1999plastid}. We take $\tau=4$ $\mu m$. 

The expression for $d$ requires $d^*-k_A>0$; this provides an acceptable range for choosing $k_A<0.55$. Since $k_A$ and $k_E$ represent the degradation of the autoinducer (inside and outside the chloroplasts, respectively), we assume they are of the same order of magnitude. In addition, we need to choose $\tilde{\delta}$ such that $S\tilde{\delta}>>k_E$; this is the regime in which the autoinducer can collect to a sufficient density outside the chloroplasts to trigger the switch in $d$. Final specification of $k_A$, $k_E$ and $\tilde{\delta}$ is subject to a rough fit of the observed slope of the data in \cite{pyke1992chloroplast} to the leading order expression for $\alpha$ for each geometry. We choose to fix $k_E=0.1$ and $\tilde{\delta}=5$ for all geometries, and alter $k_A$ to adjust the slope for each geometry. There is some flexibility in choosing these parameters within an order of magnitude or more.

%
%

\begin{acknowledgements}
K Paton thanks NSERC and UBC for their financial support. E Cytrynbaum was supported by an NSERC Discovery Grant 298313-09.
\end{acknowledgements}


\begin{thebibliography}{13}
\providecommand{\natexlab}[1]{#1}
\providecommand{\url}[1]{{#1}}
\providecommand{\urlprefix}{URL }
\expandafter\ifx\csname urlstyle\endcsname\relax
  \providecommand{\doi}[1]{DOI~\discretionary{}{}{}#1}\else
  \providecommand{\doi}{DOI~\discretionary{}{}{}\begingroup
  \urlstyle{rm}\Url}\fi
\providecommand{\eprint}[2][]{\url{#2}}

\bibitem[{Briggs and Christie(2002)}]{briggs2002}
Briggs WR, Christie JM (2002) Phototropins 1 and 2: versatile plant blue-light
  receptors. Trends in plant science 7(5):204--210

\bibitem[{Dockery and Keener(2001)}]{dockery2001}
Dockery JD, Keener JP (2001) A mathematical model for quorum sensing in
  pseudomonas aeruginosa. Bulletin of mathematical biology 63(1):95--116

\bibitem[{Ellis and Leech(1985)}]{ellis1985}
Ellis J, Leech R (1985) Cell size and chloroplast size in relation to
  chloroplast replication in light-grown wheat leaves. Planta 165(1):120--125

\bibitem[{Hall and Langdale(1996)}]{hall1996molecular}
Hall LN, Langdale JA (1996) Molecular genetics of cellular differentiation in
  leaves. New phytologist 132(4):533--553

\bibitem[{Martin and Herrmann(1998)}]{martin1998gene}
Martin W, Herrmann RG (1998) Gene transfer from organelles to the nucleus: how
  much, what happens, and why? Plant Physiology 118(1):9--17

\bibitem[{Osteryoung et~al(1998)Osteryoung, Stokes, Rutherford, Percival, and
  Lee}]{osteryoung1998chloroplast}
Osteryoung KW, Stokes KD, Rutherford SM, Percival AL, Lee WY (1998) Chloroplast
  division in higher plants requires members of two functionally divergent gene
  families with homology to bacterial ftsz. The Plant Cell Online
  10(12):1991--2004

\bibitem[{Pyke(2009)}]{pykeplastidbook}
Pyke K (2009) Plastid Biology. Cambridge University Press

\bibitem[{Pyke(1999)}]{pyke1999plastid}
Pyke KA (1999) Plastid division and development. The Plant Cell Online
  11(4):549--556

\bibitem[{Pyke and Leech(1991)}]{pyke1991rapid}
Pyke KA, Leech RM (1991) Rapid image analysis screening procedure for
  identifying chloroplast number mutants in mesophyll cells of arabidopsis
  thaliana (l.) heynh. Plant Physiology 96(4):1193--1195

\bibitem[{Pyke and Leech(1992)}]{pyke1992chloroplast}
Pyke KA, Leech RM (1992) Chloroplast division and expansion is radically
  altered by nuclear mutations in arabidopsis thaliana. Plant physiology
  99(3):1005--1008

\bibitem[{Pyke and Leech(1994)}]{pyke1994genetic}
Pyke KA, Leech RM (1994) A genetic analysis of chloroplast division and
  expansion in arabidopsis thaliana. Plant Physiology 104(1):201--207

\bibitem[{Pyke et~al(1994)Pyke, Rutherford, Robertson, and
  Leech}]{pyke1994arc6}
Pyke KA, Rutherford SM, Robertson EJ, Leech RM (1994) arc6, a fertile
  arabidopsis mutant with only two mesophyll cell chloroplasts. Plant
  Physiology 106(3):1169--1177

\bibitem[{Van~Delden and Iglewski(1998)}]{van1998cell}
Van~Delden C, Iglewski BH (1998) Cell-to-cell signaling and pseudomonas
  aeruginosa infections. Emerging infectious diseases 4(4):551

\end{thebibliography}

%

\end{document}